\documentstyle[multicol,aps,floats,epsfig]{revtex}
 
\begin{document}
\draft

\twocolumn[\hsize\textwidth\columnwidth\hsize\csname@twocolumnfalse\endcsname

\title{Percolative conductivity and critical exponents 
in mixed-valent manganites}

\author{Ye Xiong$^{1}$, Shun-Qing Shen$^{2,3}$, and X.C. Xie$^{1}$}

\address{$^{1}$Department of Physics, Oklahoma State University, Stillwater,
Oklahoma 74078}

\address{$^{2}$Department of Physics,The University of Hong Kong, 
Pokfulam, Hong Kong, China}

\address{$^{3}$Institute of Physics, Chinese Academy of Sciences, P. O. Box 603, Beijing 100080, China}
\date{\today}

\maketitle
\begin{abstract}
Recent experiments
have shown that some colossal magnetoresistance (CMR) 
materials exhibit a percolation transition.
The conductivity exponent varies substantially
with or without an external magnetic field. This
finding prompted us to carry out theoretical studies of
percolation transition in CMR systems.
We find that the percolation transition coincides with the
magnetic transition and this causes a large effect of a magnetic
field on the percolation transition.
Using real-space-renormalization method and numerical calculations
for two-dimensional (2D) and three-dimensional (3D) models,
we obtain the conductivity exponent
$t$ to be 5.3 (3D) and 3.3 (2D) without a magnetic field, and
1.7 (3D) and 1.4 (2D) with a magnetic field.
\end{abstract}


\bigskip]

Colossal magnetoresistance (CMR), an unusual large change of resistivity in
the presence of a magnetic field, has been extensively studied in
ferromagnetic perovskite manganites.\cite{Tokura00} It is well known that
the electronic phase diagrams of CMR materials are very complex. There are
various ordering states and phase transitions as the carrier concentration
is varied. For example, La$_{1-x}$Ca$_{x}$MnO$_{3}$, a typical double
exchange ferromagnetic metal when x=0.33, becomes a charge ordered insulator
with a specific type of electronic orbital and magnetic orderings when
x=1/2. \cite{Schiffer95} Recently, it was demonstrated that La$_{1-x-y}$Pr$%
_{y}$Ca$_{x}$MnO$_{3}$ (x=3/8) system, where Pr is chosen to vary the
chemical pressure, may be electronically phase separated into a
sub-micrometer-scale mixture of insulating regions and ferromagnetic (FM)
metallic regions.\cite{Uehara99,Babu} Experimental findings of FM clusters
and phase separation in CMR materials have also been reported in early
studies.\cite{Zhao} 
Electron diffraction and dark-field imaging on the 
La$_{1-x-y}$Pr$_{y}$Ca$_{x}$MnO$_{3}$ samples
indicate that the insulating region is a x=1/2 charge ordered phase. This is
not a charge congregation type of phase separation, which was observed in
slightly doped antiferromagnetic manganites \cite{Allodi97} and was
extensively discussed. \cite{Moreo99} CMR effect was observed in different
samples with $0.275<y<0.41$, and was explained by percolative transport
through the ferromagnetic domains. According to the percolation theory, the
conductivity $\sigma \propto (p-p_{0})^{t}$, where $p$ is
the concentration of the metallic phase
and $p_{0}$ is its critical value. In Ref.\cite{Uehara99},
the exponent was studied in the presence and absence of an
external magnetic field and two values are substantially different.
Thus, the experiment shows that the percolative transport in the CMR
systems depends sensitively on the relative spin orientation of adjacent
ferromagnetic domains which is controlled by an applied magnetic
field. The goal of
this work is to develop a percolation theory which takes into account the
magnetic transition in the CMR materials. In particular, we would like to
understand what causes the exponent to be so different with or without
an external magnetic field.

Based on phenomenological considerations, we study two- or three-dimensional
(2D or 3D) lattice percolation models. The conductivity and its critical
exponent are calculated by means of real space renormalization and numerical
methods. The system consists of three types of lattice sites. Each site has
spin $S_{i}(=\pm 1,0)$ where $i$ denotes the index of a site. This model is
similar to the site-diluted spin system used in Ref. \cite{Marinari}. $%
S_{i}=0$ implies that the site is empty, meaning occupied by the x=1/2
charge ordered phase. $S_{i}=\pm 1$ means the site is occupied by the
ferromagnetic metallic phase with up and down magnetizations.\cite{note1}
The Hamiltonian of the spin interaction is Ising-like and is written as 
\begin{equation}
H_{s}=-J\sum_{\langle i,j\rangle }S_{i}\cdot S_{j}+\sum_{i}H\cdot S_{i}.
\end{equation}
Here $\langle i,j\rangle $ denotes a pair of the nearest neighbor sites, $J$
is the interaction energy, and $H$ is the strength of the external magnetic
field. We consider a ferromagnetic interaction, i.e., $J>0$. In order
to investigate transport properties, we assume that the local conductivity
between two nearest occupied sites is either $0$ for antiparallel spins ($%
\uparrow \downarrow $ and $\downarrow \uparrow $) or $1$ for parallel spins (%
$\uparrow \uparrow $ and $\downarrow \downarrow $) and is zero if one or both
sites are empty. Thus, the conductivity between neighboring sites can be
expressed as follows 
\begin{equation}
\sigma _{ij}=\left\{ 
\begin{array}{ll}
1, & S_{i}\cdot S_{j}=1; \\ 
0, & \text{otherwise. }
\end{array}
\right.  \label{locon}
\end{equation}

First, we use the standard Monte-Carlo method to study the magnetic
properties of the model. Knowing the spin structure is necessary for the
transport studies since the local conductivity depends on the spins through
Eq.(2). The conductance calculation is performed on a set of spin
configurations produced by the Markov chain. In the Markov chain every spin
configuration is generated from the previous one by using the probability $%
e^{-\beta \Delta H_{s}}/(e^{-\beta \Delta H_{s}}+1)$, where $\Delta H_{s}$
is the energy difference between these two configurations and $\beta =1/
k_{B}T$.\cite{monte} The calculations are carried out on finite square and
cubic lattices for 2D and 3D systems and the periodic boundary condition is
adopted to eliminate the boundary effects. The conductance $G$ for every
spin configuration in the Markov chain is obtained by calculating the total
conductance of the resistor network.\cite{Derrida} In the resistor network,
the local conductivity between neighboring sites is determined by Eq.(2).
The total conductance is calculated for many samples to obtain the average
conductance.

When the charge order (CO) phase is dominant in some of the CMR materials,
most sites are empty according to our model. Thus, most of the conductivities
between neighboring sites are zero, corresponding to the low concentration
limit ($p\sim 0$) in the percolation model, where $p$ is the probability of
non-zero local conductivity. The small FM islands is well separated by the
CO phase. Because the spin correlations between FM blocks are cut off by the
CO phase, the spin orientation for each FM block is random, either up or
down. Therefore, the spontaneous magnetization $m$ will be zero at any
temperature $T$. As $p$ increases, $m$ continues to be zero until $p$
reaches $p_{0}$ at which the first infinite FM cluster appears. 
If $p>p_{0}$,
a finite spontaneous magnetization appears for $T<T_{c}(p)$ with $T_{c}(p)$
the critical temperature at concentration $p$. In Figs. 1(a) and 1(b) we
plot the normalized magnetization $m $ as a function of temperature $T$ with
different concentration probabilities $p$ for 2D and 3D systems, respectively,
calculated using the Monte Carlo method. The interaction strength $J$ is set
to unity. It can be seen that when the temperature increases $m$ is reduced
and reaches zero at $T_{c}(p)$. From magnetization data with
different size samples and through finite-size scaling,
we can determine the
critical temperature. In Figs. 2(a) and 2(b) we plot the relation between
the critical temperature $T_{c}(p)$ and the concentration $p$. As shown in
these plots, $T_{c}(p)$ will approach the Curie temperature $T_{c}$ for the
regular Ising model as $p$ approaches to $1$. We find that $T_{c}$ 
determined from Fig.2 is $2.3J$
or $4.4J$ for 2D or 3D system. Both of them are in good agreement with
the values of the regular Ising model.\cite{Plischke}

\begin{figure}[tbp]
\centerline{\epsfig{file=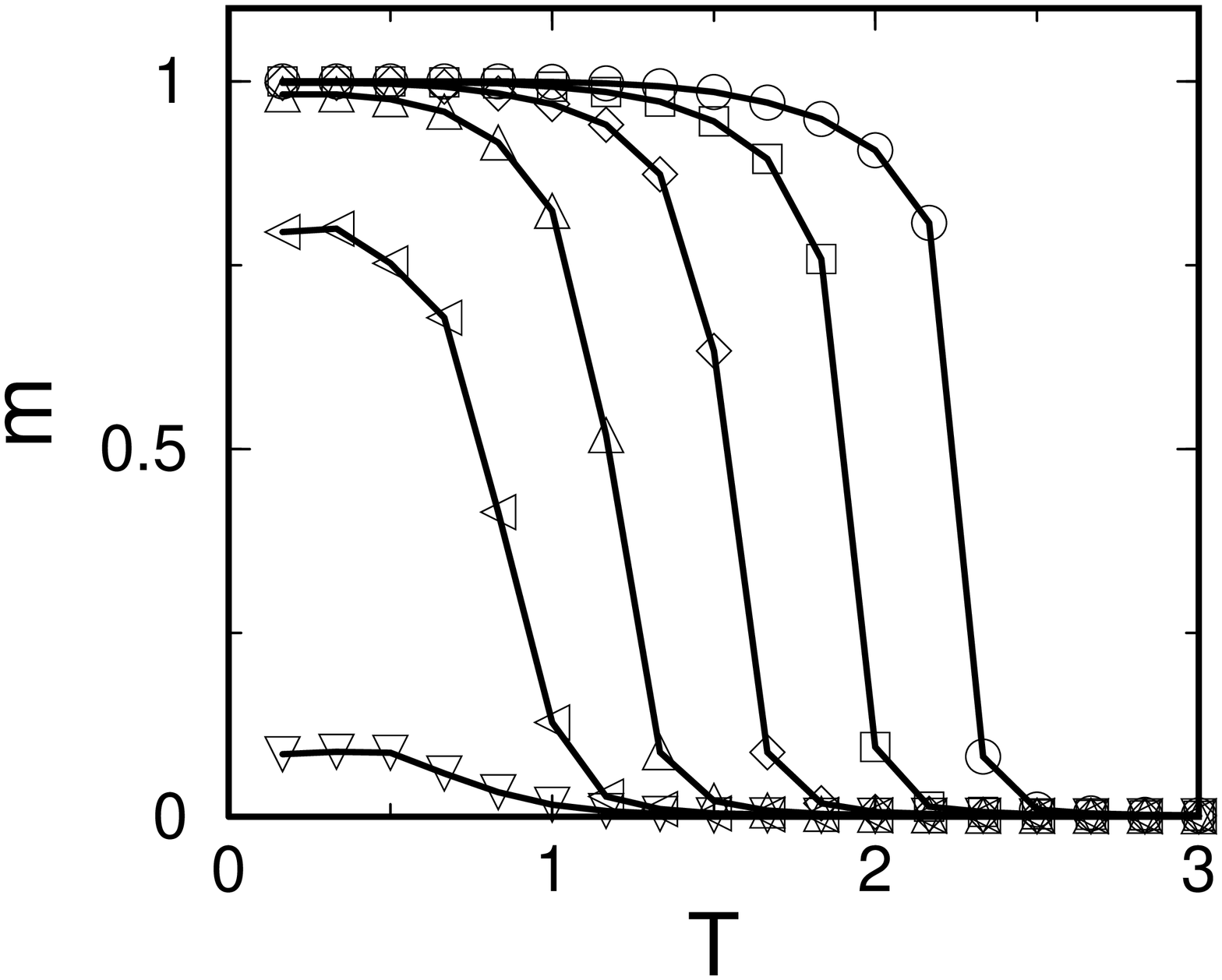, width=7cm, angle=270}}
\vspace{0.5cm}
\centerline{\epsfig{file=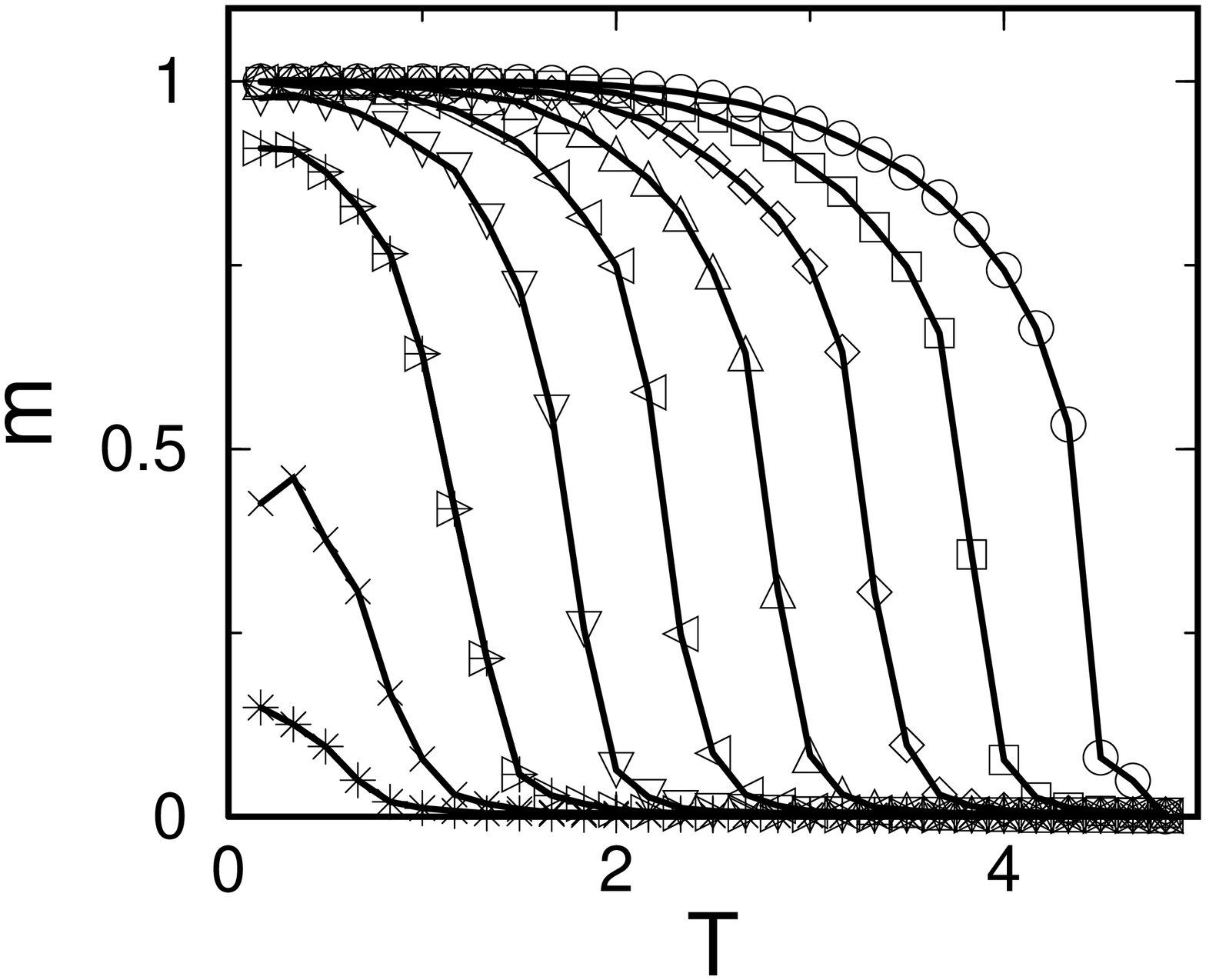, width=7cm, angle=270}}
\vspace{0.7cm}
\caption{ (a)The normalized spontaneous magnetization $m $ as a
function of temperature $T$ for different concentration $p$ in the 2D model.
The unit of $T$ is set by the interaction strength $J/k_{B}$.
From top to bottom, the concentrations are $p=1.0$, $0.9$, $0.8$, $0.7$, $%
0.6 $ and $0.5$ The
calculation is performed on $1000$ samples whose size is $100 \times 100$.
(b) $m $ as a function of $T$ in the 3D model. The calculation is
performed on $200$ samples whose size is $10 \times 10 \times 10$. From top
to bottom, the concentrations are $p=1.0$, $0.9$, $0.8$, $0.7$, $0.6$, $0.5$%
, $0.4$, $0.3$ and $0.2$. }
\end{figure}

\begin{figure}[tbp]
\epsfig{file=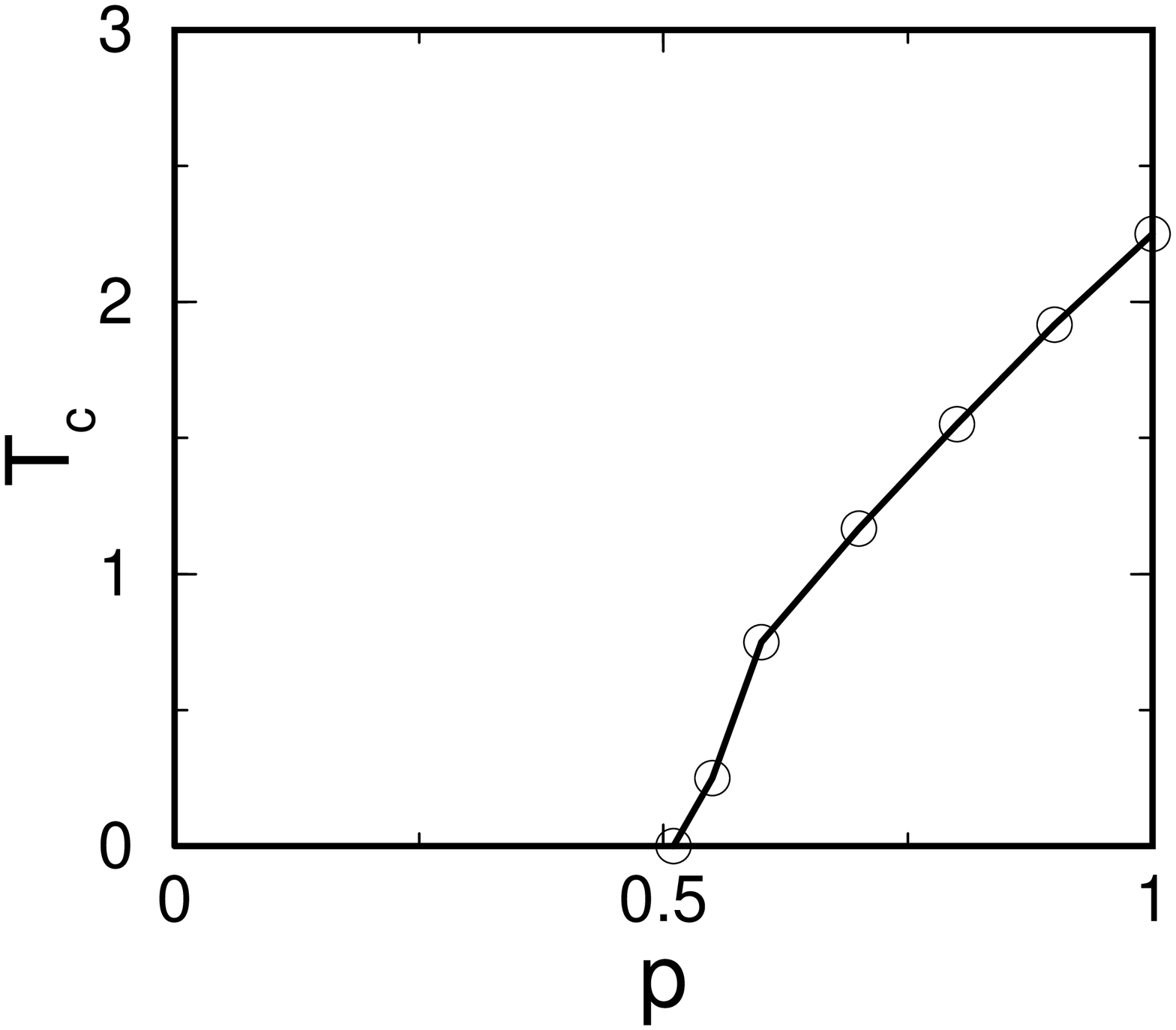, width=3.1cm, angle=270}
\epsfig{file=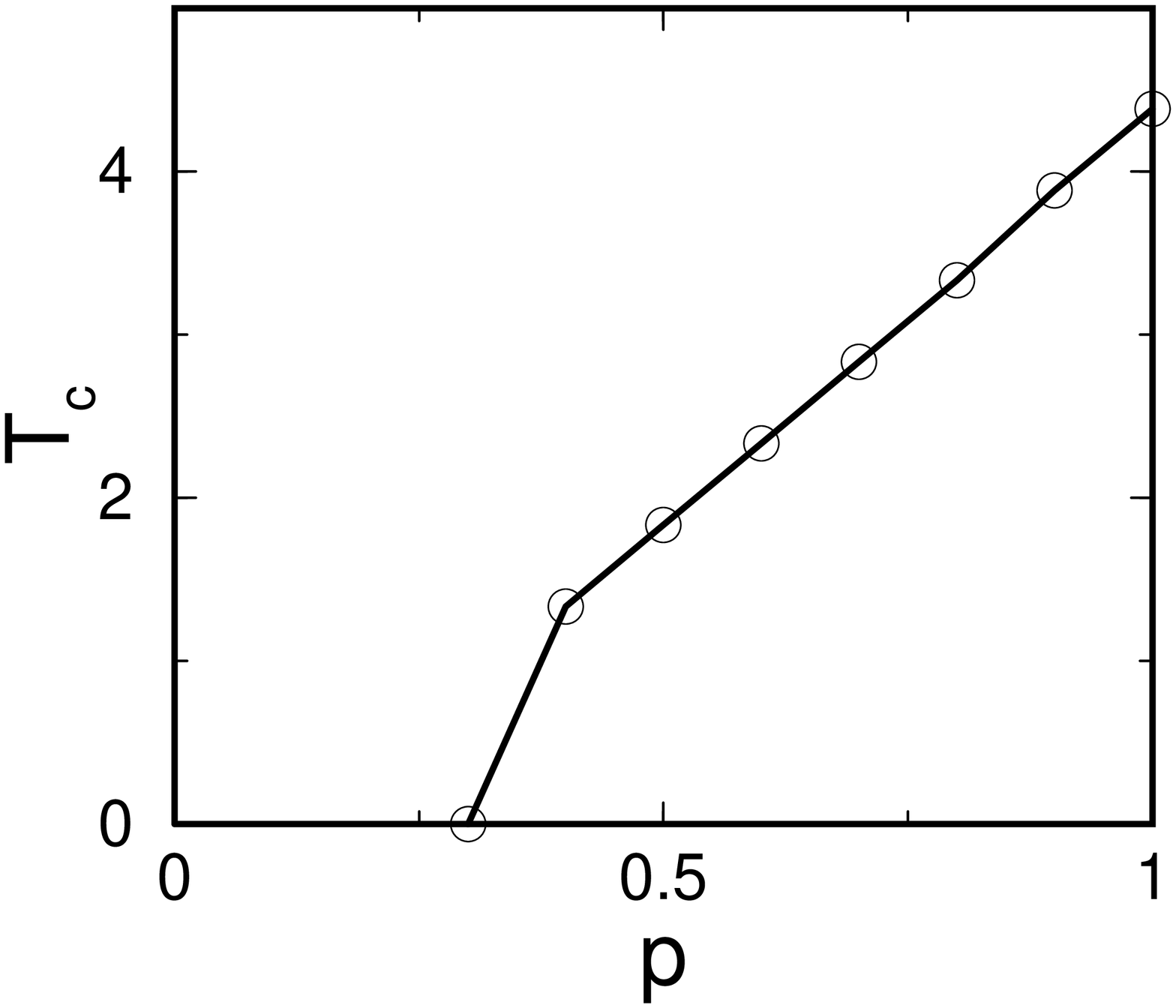, width=3.1cm, angle=270}
\vspace{0.7cm}
\caption{ (a) The critical temperature $T_c$ as a function of the
concentration $p$ in 2D. 
(b) The critical temperature $T_c$ as a function of the concentration 
$p $ in 3D. }
\end{figure}

From the spin configurations of the system, we can calculate the conductance
by using the local conductivity defined in Eq. (\ref{locon}). In a
paramagnetic phase, there is no infinite cluster with spins pointing in the
same direction. Hence, according to Eq. (\ref{locon}) there is no conducting
path throughout the sample. On the other hand, the first conducting path
appears simultaneously when the magnetization starts becoming non-zero. 
{\it Thus, the phase transition from the FM phase to the paramagnetic phase is
accompanied by the metal-insulator (MI) transition in the conductance.} This
implies that at zero temperature the MI transition occurs at the critical
concentration of the percolation threshold $p_{0}$. Near the critical point
the averaged conductance $G$ can be expressed as $G\sim (p-p_{0})^{t}$,
where $t$ is the the conductance critical exponent for the transition. 

\begin{figure}[tbp]
\centerline{\epsfig{file=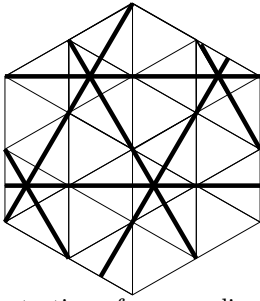, width=3.5cm, angle=0}}
\caption{ To illustration of renormalization in 2d triangle lattice }
\end{figure}

We now discuss the real space renormalization method. We start with a 2D
system and later extend to a more complicated 3D system. The essential
physics does not depend on dimensionality, although numerical numbers do.
Consider a triangular lattice in a 2D plane. The choice of a triangular
lattice is for the convenience of rescaling of the system in the real space
renormalization procedure. By enlarging the system by factor $\sqrt{3}$ and
grouping every three sites into a ''supersite'', the number of the
supersites is the same as the number of sites in the original system (see
Fig.3). In Fig.3, the thin lines are for the original system and the thick
lines are for the rescaled system. A supersite is regarded as empty if the
majority of the three sites are empty. Thus the concentration of the new
system can be expressed as 
$p^{\prime }=p^{3}+3p^{2}(1-p)$(Ref. \cite{percolation}).
Near the critical point $p_{0}=0.5$, on the metallic side, $G$ can be
written as $G=G_{0}(p-p_{0})^{t}$, where the constant $G_{0}$ is
proportional to the conductance of the unit cell. For the enlarged system
this relation becomes $G^{\prime }=G_{0}^{\prime }(p^{\prime }-p_{0})^{t}$.
In a 2D system the conductance is independent of the system size and we have 
$G=G^{\prime }$. Thus, $t=\ln (\frac{G_{0}}{G_{0}^{\prime }})/\ln (\frac{%
p^{\prime }-p_{0}}{p-p_{0}})$. If the spin degree of freedom is frozen, the
conductance of elementary cell is inversely proportional to the size of the
cell, hence, $t=\ln (\sqrt{3})/\ln (\frac{3}{2})=1.35$. This value is very
close to the exact value of $t=4/3$ in the standard 2D percolation model. 
\cite{Efros} When taking into account the spin degrees of freedom, as we
discussed above, the $p_{0}$ point is also the critical point for the
spontaneous magnetization. The conductance $G_{0}$ (or $G_{0}^{\prime }$) is
now associated with spin configurations. 
The spin dependence comes from the fact that $G_{0}$ is
proportional to the average of conductivity between two nearest neighbor
sites which is spin dependent, 
\begin{eqnarray}
G_{0} &\sim &\langle \sigma _{ij}\rangle \sim \left\langle \cos (\frac{%
\theta _{i}-\theta _{j}}{2})\right\rangle  \nonumber \\
&=&\frac{1}{A}\int\limits_{0}^{\pi }{d\theta _{i}}\int\limits_{0}^{2\pi }{%
d\phi _{i}}\int\limits_{0}^{\pi }{d\theta _{j}}\int\limits_{0}^{2\pi }{%
d\phi _{j}}  \nonumber \\
&&\times e^{-\beta mS\cos \theta _{i}}e^{-\beta mS\cos \theta _{j}}\cos (%
\frac{\theta _{ij}}{2})
\end{eqnarray}

The conductivity expression between nearest neighbor spin $\sigma _{ij}\sim
\cos (\frac{\theta _{ij}}{2})$ comes from the double-exchange model where $%
\theta _{ij}$ is the angular difference between spins $S_{i}$ and $S_{j}$
,which satisfies\cite{doubleexchang} 
\begin{equation}
\cos \theta _{ij}=\cos \theta _{i}\cos \theta _{j}+\sin \theta _{i}\sin
\theta _{j}\cos \left( \phi _{i}-\phi _{j}\right)
\end{equation}
$(\theta _{i},\phi _{i})$ denotes the orientation of spin $S_{i}$. A is a
normalization constant. From the above equation, it is easy to show that $%
G_{0}\sim m^{2}$. Because $m$ can be written as $m\sim (p-p_{0})$,\cite
{percolation} we finally get the conductance critical exponent in 2D
triangular lattice as $t^{\prime }\simeq 1.35+2=3.35$.

In the 3D case, we consider the normal cubic lattice. The elementary vectors
of the enlarged lattice is just two times of those of the original one $\{ 
\hat{e_x}^{\prime} , \hat{e_y}^{\prime} , \hat{e_z}^{\prime} \}=\{2\hat{e_x}
, 2\hat{e_y} , 2\hat{e_z} \} $. In this case $p^{\prime} =\sum_{ n=4 }^{8}
C_{8}^{n}p^n (1-p)^{8-n} $, from which we conclude that $p_0=0.395 $ and $%
t=1.7 $ without spin effects. After the spin degrees of freedom are
considered, the formula $G_{0} \sim m^2$ is still satisfied. But in 3D case $%
m$ has $m \sim (p-p_0)^{1.79} $.\cite{percolation} 
So the critical exponent 
$t^{\prime} \simeq 1.7+2 \times 1.79 \simeq 5.3$.

In Fig. 4 we show the numerical results of 2D conductance $G$. 3D
calculation has not been done because of computational limit. The points
with steepest drop are defined as the critical points for the
metal-insulator transition. These points are consistent with the magnetic
critical points (see Fig.1), as we discussed before. In the inset of Fig.4
we show the $\ln (G)$ versus $\ln (p-p_{0})$. They exhibit a linear
dependence and the slope of the curve, corresponding to the exponent $t$, 
is roughly $3.5$. This value is in a good agreement with the
estimation from the renormalization group consideration although it is
obtained from a different type of lattice structure. In Table I we list the
results of this work and compare them with the previous studies of the
standard percolation theory\cite{Efros} and experimental exponents in CMR
materials.\cite{Uehara99,Babu} The experimental exponent $t_{3}$ (or $%
t_{3}^{\prime }$) was obtained in the absence (or presence) of an external
magnetic field.

\begin{figure}[tbp]
\centerline{\epsfig{file=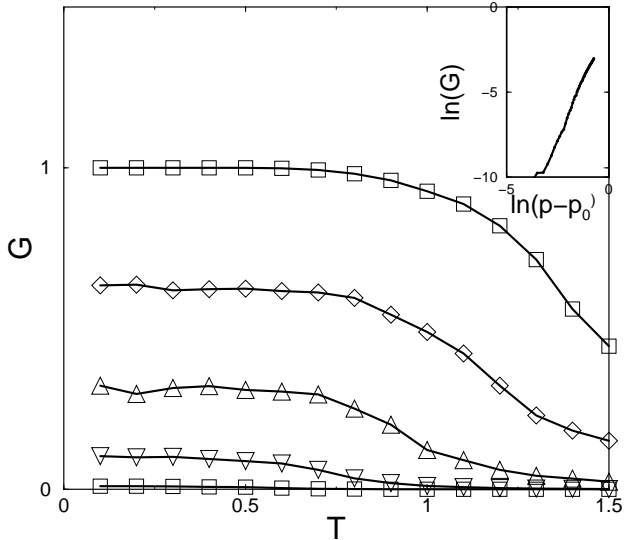, width=7.5cm, angle=270}}
\caption{ The normalized conductivity $G$ as a function of temperature $%
T $ in 2D model. From top to bottom, the concentrations are $p=1.0, 0.9,
0.7, 0.6 $ and $0.5 $. Inset: $log G$ vs. $log(p-p_0) $ shows a
linear behavior with a slope $\simeq 3.5$ }
\end{figure}

\begin{table}[!htb]
\caption{The critical exponent $t$ for 2d and 3d cases. $p_{c2}$ and $p_{c3}$
are percolation threshold for 2d and 3d, respectively. $t_{d}$ and
$t^{\prime}_{d} $ are the critical exponents for d-dimension model without or
with spin effects. Previous results are from Ref.14.
Experiment I is from Ref.3 and Experiment II is
from Ref.4.
}
\label{tab;stdev}\centering 
\begin{tabular}{ccccccc}
& $p_{c2}$ & $p_{c3}$ & $t_{2}$ & $t_3$ & $t^{\prime}_2$ & $t^{\prime}_3 $
\\ 
Previous results & $0.5$ & $0.31$ & $1.33$ & $1.9$ & - & - \\ 
Experiment I & - & - & - & 2.6 & - & 6.1 \\ 
Experiment II & - & - & -& - & - & 2.6 \\ 
Renormalization method & $0.5$ & $0.37$ & $1.35$ & $1.7$ & $3.3$ & $5.3$ \\ 
Numerical results & 0.51 & 0.32 & 1.30 & - & 3.5 & -
\end{tabular}
\end{table}

Before summary, we would like to make couple comments. (i) In order to 
calculate the conductivity exponent, we have to determine the dependence 
of $G_0$ (see Eq.(3)) on the magnetization $m$,
and the double-exchange model
is used to achieve that goal. However, the double-exchange may not
be the origin of the ferromagnetism in doped manganites,
as shown in a recent work.\cite{Zhao1} This might explain some of
the discrepancy
between the experimental and our theoretical values in the
conductivity exponent. We should also mention that the values from
the two experimental groups are not in good agreement with each other as seen
in Table I. (ii) We have neglected the quantum effects in this work. 
This might be justified because $T_c$ is relatively high in these samples. 
Develop a semi-classical transport theory in this problem is a difficult
task because of the finite phase coherence length. There are 
attempts\cite{Shi} of using the semi-classical theory to understand
the two-dimensional metal-insulator transition.

In summary, we argue that the percolation threshold corresponds
not only to the appearance of an infinite metal cluster, but also
to the phase transition from PM to FM. This coincidence of two
phase transitions renormalizes the critical exponents.
The conductivity exponent
has been obtained using real space renormalization
and numerical calculations. The exponent is found to be quite different
whether the magnetic transition is considered. This finding explains the
large exponent discrepancy in some of the
CMR materials in the presence or absence
of an external magnetic field. 

\bigskip

This work was supported by DOE and a RGC grant of Hong Kong.
We thank Junren Shi for many helpful discussions.


\begin{references}
\bibitem{Tokura00}  See {\it Colossal magnetoresistive Oxides}, ed. by Y.
Tokura, (Gordon and Breach Sciences Publishers, 2000).

\bibitem{Schiffer95}  P. Schiffer, A. P. Ramirez, W. Bao, and S. W. Cheong, 
{\it Phys. Rev. Lett.} {\bf 75}, 5144 (1995).

\bibitem{Uehara99}  M. Uehara, S. Mori, C. H. Chen, and S. W. Cheong, {\it %
Nature(London)} {\bf 399}, 560 (1999).

\bibitem{Babu} N.A. Babushkina {\it et al. Phys. Rev. B} {\bf 62},
R6081 (2000).

\bibitem{Zhao} G.M. Zhao, H. Keller, J. Hofer, A. Shengelaya, and
K.A. M\"{u}ller, {\it Solid State Commun.} {\bf 104}, 57 (1997);
M.R. Ibarra {\it et al., Phys. Rev. B} {\bf 57}, 7446 (1998).

\bibitem{Allodi97}  G. Allodi {\it et al., Phys. Rev. B} {\bf 56}, 6036
(1997); Hennion {\it et al., Phys. Rev. Lett.}{\bf \ 81}, 1957 (1998).

\bibitem{Moreo99}  A. Moreo, S. Yunoki and E. Dagotto, {\it Science}{\bf \
283}, 2034 (1999) and references therein.

\bibitem{Marinari}  E. Marinari {\it et al.}, {\it Phys. Rev. B {\bf 62},}
4999 (2000).

\bibitem{note1}  To simplify our problem, we just take two possible
orientations for magnetization of ferromagnetic domains in the mixtures.

\bibitem{Derrida}  B. Derrida and J. Vannimenus {\it J. Phys. A: Math. Gen.}
{\bf 15}, L557-L564 (1982). 

\bibitem{monte}  M. Suzuki, {\it Quantum Monte Carlo Methods in Condensed
Matter Physics}, (World Scientific Publishing Co. Pte. Ltd.).

\bibitem{Plischke}  See, for example, M. Plischke and B. Bergersen, {\it %
Equilibrium Statistical Physics}, (World Scientific, 1989).

\bibitem{percolation}  A. Bunde and S. Havlin, {\it Fractal and disordered
systems} (Springer-Verlag Berlin Heidelberg New York 1996).

\bibitem{Efros}  B. I. Shklovskii and A. L. Efros, {\it Electronic properties
of doped semiconductors}, (Springer-Verlag Berlin Heidelberg New York Tokyo
1984).

\bibitem{doubleexchang}  See, for example, E. L. Nagaev, 
{\it Physics of Magnetic Semiconductors},
(Moscow, Mir Publ, 1979). 

\bibitem{Zhao1} G.M. Zhao, {\it Phys. Rev. B}{\bf 62}, 11639 (2000).

\bibitem{Shi} J.R. Shi and X.C. Xie, {\it Phys. Rev. B}{\bf 63},
045123 (2001); J.R. Shi, S. He, and X.C. Xie, cond-matt/9904393. 
\end{references}
\end{document}